\documentstyle[twoside,epsfig,psfig]{article} 
 
\input ibvs2.sty 
 
\begin{document} 
\IBVShead{5932}{26 March 2010} 
 
\IBVStitle{BVR$_{\rm C}$I$_{\rm C}$ photometric evolution of the very fast}
\vskip -0.3 cm
\IBVStitle{Nova Ophiuchi 2010 N.1 = V2673 Oph}
\IBVStitle{} 
  
\IBVSauth{U. Munari$^1$, S. Dallaporta$^2$}

\IBVSinst{INAF Osservatorio Astronomico di Padova, Sede di Asiago, I-36032 Asiago (VI), Italy} 
\IBVSinst{ANS Collaboration, c/o Astronomical Observatory, 36012 Asiago (VI), Italy} 
 
\IBVStyp{Nova} 
\IBVSkey{photometry} 
\IBVSabs{The very fast, FeII-class Nova Ophiuchi 2010 N.1 rised rapidly to maximum
brightness, that was reached on Jan. 18.3, 2010 at $V$=8.5, $B$$-$$V$=+0.95, 
$V$$-$$R_{\rm C}$=+0.75, and  $V$$-$$I_{\rm C}$=+1.50. The rapid and smooth
decline was charaterized by $t_2^{V}$=10.0 and $t_3^{V}$=23.5 days. The
reddening affecting the nova is $E_{B-V}$=0.7 and its distance $d$=7.4~kpc,
for an height above the galactic plane of $z$=0.6~kpc.}

\begintext 
Nova Ophiuchi 2010 N.1 (= V2673 Oph) was discovered by H. Nishimura on Jan.
15.9 UT (cf. Nakano 2010) and confirmed spectroscopically by H. Maehara (2010)
as a "Fe II" class nova.  

We obtained $B$$V$$R_{\rm C}$$I_{\rm C}$ photometry of Nova Ophiuchi 2010
N.1 with a 0.30-m Meade RCX-400 f/8 Schmidt-Cassegrain telescope equipped
with a SBIG ST-9 CCD camera.  The photometry was accurately corrected for
color equations using nightly calibrations on Landolt (1992, 2009) standard
stars.  The data are presented in Table~1, and plotted in Figure~1.  The
combined (Poissonian + transformation) errors (always less than 0.03 mag) do
not exceed the dimension of the symbols in Figure~1.  The zero points of the
photometry are scaled on the nearby star TYC 6260-1846-1, for which we
adopted: $B$=11.550, $V$= 10.963, $R_{\rm C}$= 10.574 and $I_{\rm C}$=
10.222.  The $B$ and $V$ are the values recommended by AAVSO for this star,
the $R_{\rm C}$ and $I_{\rm C}$ are derived combining $B$, $V$ with
$J$,$H$,$K$ from 2MASS following the recipes by Caldwell et al.  (1993).

We started our observations immediately past maximum, and thus to
reconstruct the whole lightcurve as presented in Figure~1,  we had to 
integrate them with the published data.

Various estimates, based on unfiltered CCD observations secured around the
time of discovery with digital cameras by Japanese amateurs, were published
in CBET 2128.  These observations are generally calibrated against the
$R_{\rm C}$ band values of field stars as listed by the USNO catalog.  We
have measured the field stars around Nova Ophiuchi 2010 N.1 and found a mean
$<$$V$$-$$R_{\rm C}$$>$=+0.57 for them.  We thus applied this shift to the
unfiltered photometry of CBET 2128 and inserted it as open circles in
Figure~1.

Four aproximately $V$-band observations were obtained by Vollmann (2010)
from the green channel of color CCD images obtained with a DSLR camera. 
Comparison with our simultaneous photometry indicates that Vollmann values
need to be corrected by +0.1 mag to be placed onto the $V$ photometric
scale.  We applied such a correction and plotted the data as star symbols 
in Figure~1.

The VSNET organization collected some $B$$V$$R_{\rm C}$$I_{\rm C}$ CCD
photometric data of Nova Ophiuchi 2010 N.1, with observers S.  Kiyota and H. 
Maehara (cf March 1, 2010 summary in [vsnet-recent-nova 35402] at
http://www.kusastro.kyoto-u.ac.jp/vsnet/).  The data obtained by observer S. 
Kiyota were corrected for instrumental color equations, and are inserted in
Figure~1 as asterisks.  They did not require adjustments, as it also was for
$V$ band data by VSNET observer H.  Maehara.  The $B$,$R_{\rm C}$ and
$I_{\rm C}$ data of the latter, however, need the application of a shift to
be brought in agreement with the rest of the data.  The shift we applied
amounts to +0.32 mag in $B$, +0.34 in $R_{\rm C}$, and +0.45 mag in $I_{\rm
C}$.

\IBVSfig{18.8cm}{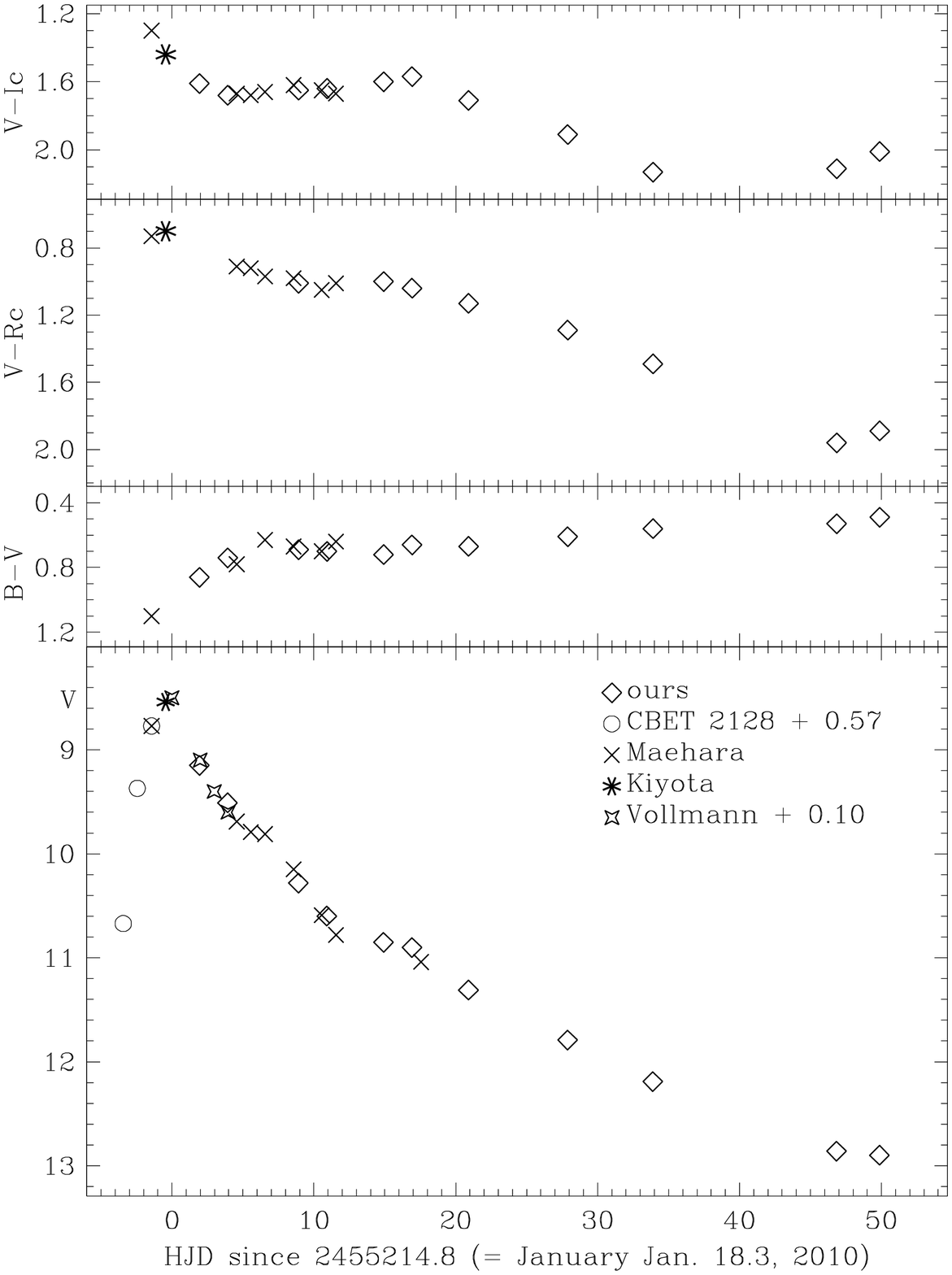}{$B$$V$$R_{\rm C}$$I_{\rm C}$ photometric
evolution of the outburst of Nova Ophiuchi 2010 N.1. For the literature
data, see text for details.}

\vskip 1cm

\centerline{Table 1. Our $B$$V$$R_{\rm C}$$I_{\rm C}$ of Nova Oph 2010 N.1}
\vskip 3mm
\begin{center}
\begin{tabular}{rrrrr}
\hline
\multicolumn{5}{c}{}\\
\multicolumn{1}{c}{HJD} &
\multicolumn{1}{c}{$V$} &
\multicolumn{1}{c}{$B$$-$$V$} &
\multicolumn{1}{c}{$V$$-$$R_{\rm C}$} &
\multicolumn{1}{c}{$V$$-$$I_{\rm C}$} \\
\multicolumn{5}{c}{}\\
  2455216.7306   &  9.15   &  +0.86     &             &  +1.61  \\
  2455218.7244   &  9.51   &  +0.74     &             &  +1.68  \\
  2455223.7142   & 10.28   &  +0.69     & +1.01       &  +1.65  \\
  2455225.7166   & 10.60   &  +0.70     &             &  +1.64  \\
  2455229.7095   & 10.85   &  +0.72     & +1.00       &  +1.60  \\
  2455231.7104   & 10.90   &  +0.66     & +1.04       &  +1.57  \\
  2455235.6959   & 11.31   &  +0.67     & +1.13       &  +1.71  \\
  2455242.6834   & 11.79   &  +0.61     & +1.29       &  +1.91  \\
  2455248.6852   & 12.19   &  +0.56     & +1.49       &  +2.13  \\
  2455261.6320   & 12.91   &  +0.53     & +1.96       &  +2.11  \\
  2455264.6625   & 12.93   &  +0.49     & +1.89       &  +2.01  \\
\multicolumn{5}{c}{}\\
\hline
\end{tabular}
\end{center}

\vskip 1.0cm

In Figure~1 the time is counted from maximum brightness that
was reached on Jan. 18.3, 2010 at $V$=8.5. At that time the colors
were $B$$-$$V$=+0.95, $V$$-$$R_{\rm C}$=+0.75, and  $V$$-$$I_{\rm C}$=+1.50.

van den Bergh and Younger (1987) derived a mean intrinsic color
$(B-V)_\circ$=$+$0.23 $\pm$0.06 for novae at the time of maximum, and
$(B-V)_\circ$=$-$0.02 $\pm$0.04 at $t_2$. Comparing with $B$$-$$V$=+0.95
at maximum and $B$$-$$V$=+0.68 at $t_2$ from Figure~1, the reddening
affecting Nova Oph 2010 N.1 is $E_{B-V}$=0.71, and the extinction
(assuming a standard $R_V$=3.1 interstellar law) is therefore 
$A_V$=2.2 mag.

The light-curve in Figure~1 is characterized by a rapid rise (the last
2.2 mag in $V$ band were covered in 3.4 days) and by a smooth decline,
regulated by the decline times
\begin{equation}
t_2^{V}=10.0   ~~~~~~~ t_3^{V}=23.5  ~{\rm days}
\end{equation}
which are the time taken by the nova to decline, in the $V$ band, by two and
three magnitudes, respectively, from maximum brightness.  These $t_2^{V}$
and $t_3^{V}$ values for Nova Oph 2010 are in the normal proportion found
for typical novae.  Given $t_2^{V}$, the Warner (1995) relation would
predict $t_3^{V}$=20.8, while Munari et al.  (2008) relation would give
$t_3^{V}$=23.1.  According to the classification of Warner (1995, his Table
5.4), a $t_2^{V}=10$ days qualifies Nova Oph  2010 N.1 to be classed among 
the very fast novae.

Published relations between the absolute magnitude and the rate of decline
generally take the form $M_{\rm max}\,=\,\alpha_n\,\log\, t_n \, + \,
\beta_n$.  Using the Cohen (1988) $V$-$t_2$ relation, the distance to the
nova is 8.3~kpc, and 7.5~kpc according to the Schmidt (1957) $V$-$t_3$
relation.

Buscombe and de Vaucouleurs (1955) suggested that all novae have the same
absolute magnitude 15 days after maximum light. The mean value of the
calibrations presented by Buscombe and de Vaucouleurs (1955), Cohen (1985),
van den Bergh and Younger (1987), van den Bergh (1988), and Capaccioli et
al. (1989) is $M_{15}^{V}$=$-$5.42$\pm$0.09, which provides a distance of
6.5 kpc to Nova Oph 2010 N.1 when compared to $V_{\rm 15}$=10.85 from 
Figure~1. Taking the mean of these three determinations, the distance to
Nova Oph 2010 N.1 is $d$=7.4~kpc. At a galactic latitude $b$=4.92 deg,
it corresponds to an height over the Galactic equatorial plane of
$z$=0.6~kpc, well within the range of heights reported by della Valle
and Livio (1998) for novae of the Fe II type.

\references 

  Buscombe, W., de Vaucouleurs, G. 1955, Obs. 75, 170

  Calddwell, J.A.R., Cousins, A.W.J., Ahlers, C.C. et al. 1993, SAAO Circ. 15, 1

  Capaccioli, M. et al. 1989, AJ 97, 1622

  Cohen J.G. 1985, ApJ 292, 90

  Cohen J.G. 1988, ASP Conf Ser 4, 114

  della Valle, M., Livio, M. 1998, ApJ 506, 818

  Landolt, A.U. 1992, AJ 104, 340

  Landolt, A.U. 2009, AJ 137, 4186

  Maehara, H. 2010, IAUC 9111

  Munari, U. et al. 2008, A\&A 492, 145

  Nakano, S. 2010, IAUC 9111

  Schmidt T. 1957, ZA 41, 182

  van den Bergh, S., Younger, P.F. 1987, A\&AS 70, 125

  van den Bergh, S. 1988, PASP 100, 8

  Vollman, W. 2010, CBET 2139

  Warner B. 1995, Cataclysmic Variable Stars, Cambridge Univ. Press  

\endreferences

\end{document}